\documentclass[preprint,review,12pt]{elsarticle}
\usepackage{amssymb}
\usepackage{amsthm}
\usepackage{subcaption}
\usepackage{hyperref}
\usepackage{lineno}
\usepackage[squaren]{SIunits}

\journal{Nuclear Physics B}
\begin{document}
\begin{frontmatter}
\title{
Investigating Neutron Scattering in a Spherical Proportional Counter: A Tabletop Experiment
}

\author{N.~Panchal\corref{cor1}\fnref{labela,labelk}}
\ead{np72@queensu.ca}
\author{L.~Balogh\fnref{labelk}}  
\author{J.-F. Caron,\fnref{labelk}} 
\author{G.~Giroux, \fnref{labela}} 
\author{P.~Gros,\fnref{labela}}

\cortext[cor1]{Corresponding author}
\affiliation[labela]{organization={Department of Physics, Engineering Physics \& Astronomy, Queen’s University},
            city={Kingston},
            postcode={K7L 3N6}, 
            state={Ontario},
            country={Canada}}

\affiliation[labelk]{organization={Department of Mechanical and Materials Engineering, Queen’s University},
            city={Kingston},
            postcode={K7L 3N6}, 
            state={Ontario},
            country={Canada}} 

\begin{abstract}
In this paper, we report on a tabletop experiment studying neutron scattering in a Spherical Proportional Counter using an Am-Be source. Systematic studies were carried out to investigate the effect of gas mixture, pressure, operating voltage, and sphere size on the drift time-rise time relationship of the signal in a spherical proportional counter. Our experimental results showed good agreement with MagBoltz simulations. These findings are a crucial step towards measuring the quenching factor in gases using a neutron beam for the New Experiments With Spheres-Gas (NEWS-G) experiment and has important implications for the development of neutron detection techniques and their potential applications in nuclear and particle physics.
\end{abstract}


\end{frontmatter}
\section{Introduction}
The search for dark matter, a ubiquitous yet elusive component of the universe, remains one of the most pressing questions in modern physics~\cite{darkmatr_1,darkmatr_2}. Among the many dark matter candidates, weakly interacting massive particles (WIMPs) \cite{WIMPS} are particularly intriguing, and detecting them remains a major focus of experimental efforts. The NEWS-G experiment \cite{newsg} is one such effort aimed at detecting WIMPs by measuring nuclear recoils in noble gases using a spherical proportional counter (SPC) detector \cite{BROSSARD2019412}, which offers high sensitivity due to its remarkably low energy threshold. However, accurate measurement of the nuclear recoil energy requires knowledge of the quenching factor (QF), which quantifies the reduction of ionization due to nuclear recoils compared to electronic recoils. 

Despite ongoing measurements of the QF by the NEWS-G group \cite{marie_vidal}, it is challenging to conduct extensive studies of the effects of various gas mixtures, operating voltages, pressures, and detector sizes on the QF in an in-beam experiment. In this context, we present a tabletop experiment that will facilitate experiments aimed at studying the effects of these parameters on the QF using a neutron source. In particular, we investigate drift and diffusion characteristics in the detector for different parameter settings and compare our results to existing simulation models.

\section{Experimental setup}
\label{}
\label{}
\begin{figure}[ht]
\centering
\includegraphics[width=0.6\textwidth]{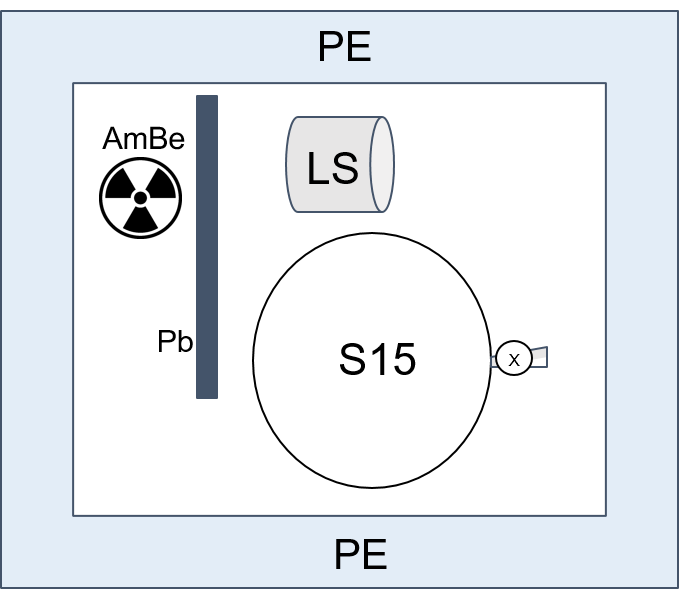}
\caption{Schematic of the set up; LS: liquid scintillator, S15: 15\,{\centi}{\meter} diameter spherical proportional counter.}
\label{fig1:schematic_setup}
\end{figure}

A schematic of the experimental setup is shown in Fig. \ref{fig1:schematic_setup}.
The SPC is made of a 15\,{\centi}{\meter} diameter aluminum sphere (S15) with a 2 mm diameter steel ball sensor \cite{Arnaud_2018},\cite{Katsioulas_2018} at the center. The wall thickness of the sphere is $\sim$\,2\,{\milli}{\meter}. For the neutrons, an Am-Be source of current activity 28\,{\milli \curie} is placed 10\,{\centi\meter} from the sphere. The neutron emission rate from the source is approximately 1000 neutrons per {\second} in all directions. Neutrons scattered from the gas inside the SPC are then detected in a backing detector (liquid scintillator, model EJ-309, Eljen \cite{eljen}) coupled to a photo-multiplier tube (R7724, Hamamatsu \cite{hamamatsu}). A 5\,{\centi\meter} thick poly-ethylene layer was placed on all sides to shield the lab environment from the neutron radiation originating from the source. Also, a $\sim$2\,{\milli\meter} thick lead sheet was placed in between the Am-Be source and the sphere to minimize $\gamma$-rays from the source.

\section{SPC}

The SPC is an innovative idea for a gaseous detector developed by I. Giomataris \textit{et.~al.} in 2006~\cite{GIOMATARIS2006208,I_Giomataris_2008}. These detectors have low intrinsic capacitance coupled with high-gain, allowing an energy threshold that is sufficiently low to detect ionization from a single electron. This makes an SPC a suitable candidate to search for low-mass dark matter for the NEWS-G experiment. In an SPC, a positive operating voltage is applied on a sensor at the center, which acts as the anode, and the sphere surface is grounded, which then becomes the cathode. The primary electrons generated in the sphere due to ionization move towards the anode due to the radial electric field. The electric field lines in the SPC are dense at the center and sparse elsewhere, which divides the detector into two field zones: drift and avalanche regions. Initially, the primary electrons drift towards the sensor. Near the sensor, the electric field becomes much stronger due to its $1/r^2$ scaling, which results in an avalanche of secondary electrons, known as a Townsend avalanche \cite{townsend}. The ions produced due to the avalanche move towards the cathode, i.e., surface of the sphere. This induces a detectable current in the sensor which results in the signal formation in the SPC. The produced signal is then fed to a charge-sensitive Canberra 2006 pre-amplifier with a time constant of the order of 100 $\mu$s.

\section{Liquid scintillator}

\begin{figure*}[ht]
    \centering
    \begin{subfigure}[b]{0.9\textwidth}
        \centering
        \includegraphics[width=0.9\linewidth]{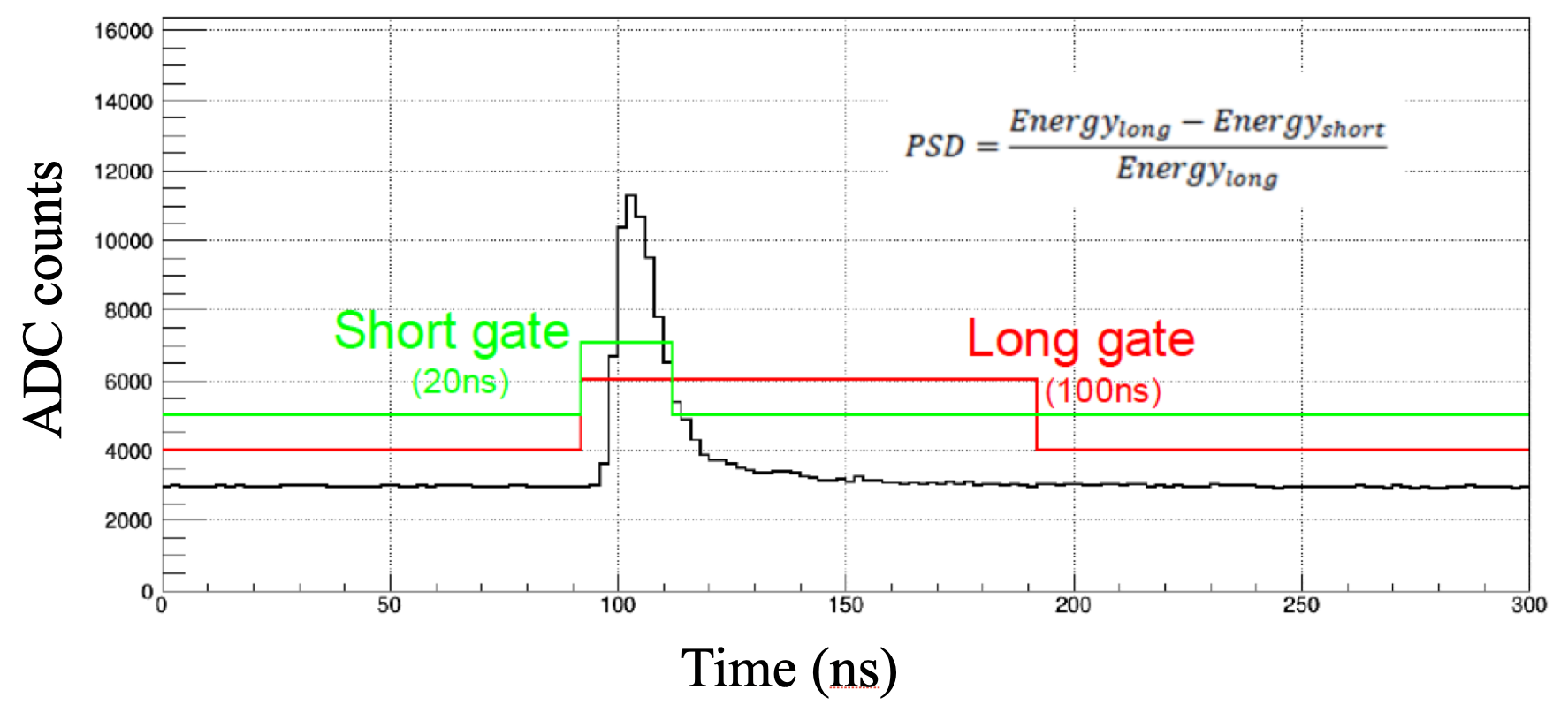}
        
        \caption{}
        \label{fig3a}
    \end{subfigure}%
  \\
    \begin{subfigure}[b]{0.9\textwidth}
        \centering
        \includegraphics[width=0.9\linewidth]{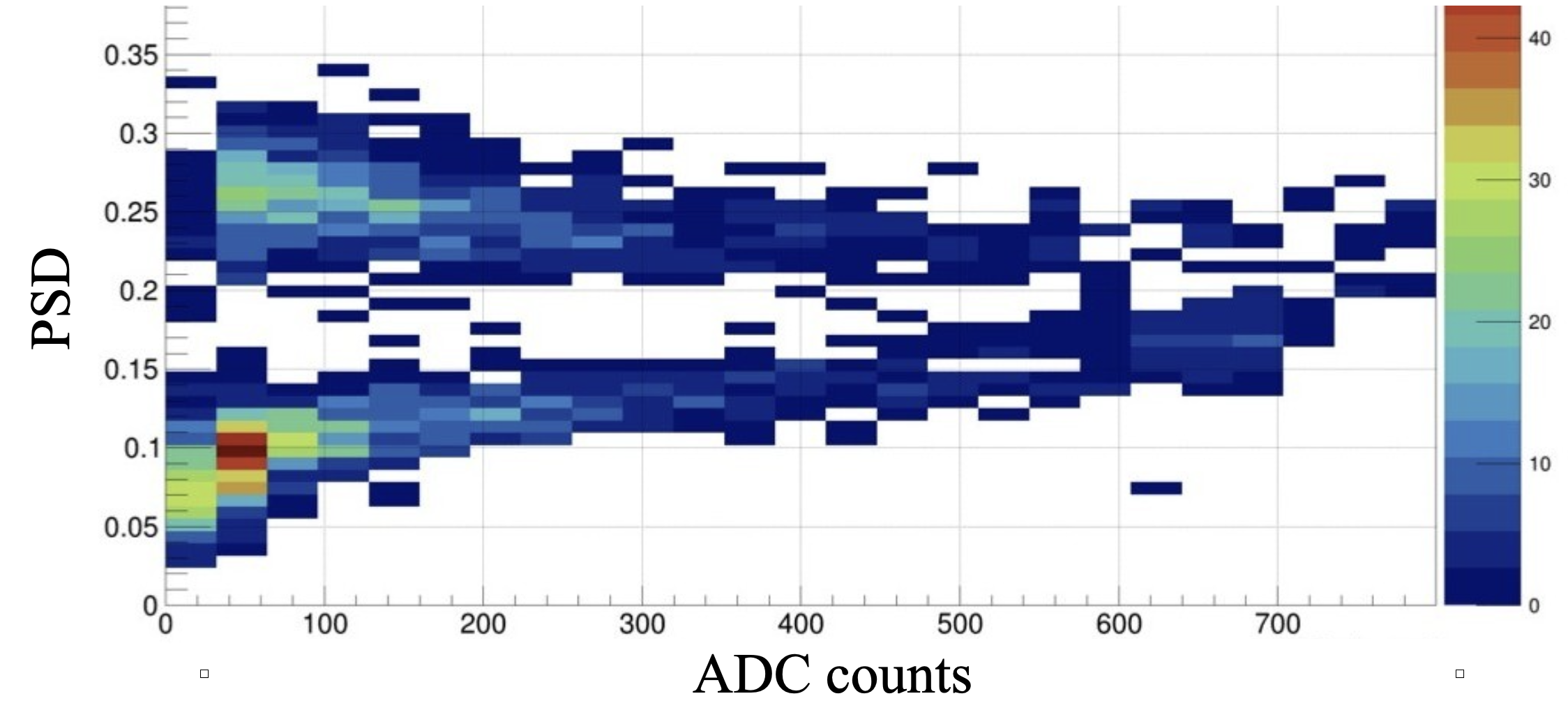}
        
        \caption{}
        \label{fig3b}
    \end{subfigure}
    \caption{(a) An example of the scintillator pulse showing long and short gates. (b) The pulse shape discrimination distribution. Here, ADC: analog-to-digital converter.}
    \label{fig3:psd_cut}
\end{figure*}

For detecting scattered neutrons from the SPC, the liquid scintillator is an ideal choice because of its fast response and excellent n/$\gamma$ discrimination \cite{Ranucci,Knoll}. The scintillation light produced in the liquid scintillator due to neutrons has a slower decay component than the one produced from $\gamma$-rays. This makes it possible to distinguish between these based on a Pulse Shape Discrimination (PSD) parameter, defined as
\begin{equation}
    PSD = \frac{Energy_{long} - Energy_{short}}{Energy_{long}}
\end{equation}
where Energy$_{long}$ and Energy$_{short}$ are the integrated charges in the long and short gates, respectively. A typical liquid scintillator pulse with long and short gates of 100 and 20 ns, respectively, is shown in Fig. \ref{fig3a}, and the corresponding PSD distribution is shown in Fig.~\ref{fig3b}, where the two different population, one for neutrons and the other for gammas can be easily seen. 

\section{Data Acquisition}
\begin{figure}[]
\centering
\includegraphics[width=0.7\textwidth]{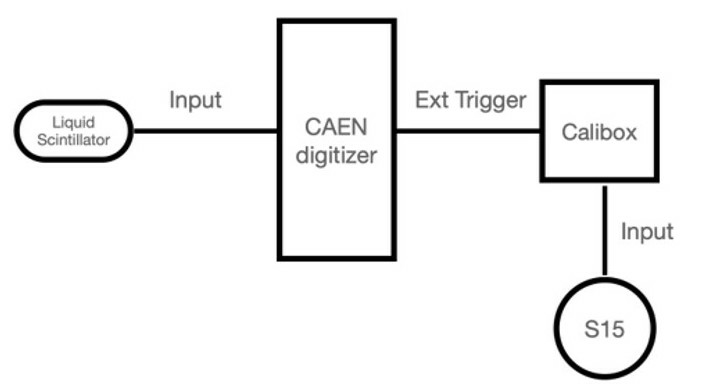}
\caption{A schematic of the data acquisition.}
\label{fig2:schematic_daq}
\end{figure}

A schematic of the data acquisition (DAQ) is shown in~Fig.~\ref{fig2:schematic_daq}. 
The output signal of the liquid scintillator is fed to a digitizer (model number 1730, CAEN). One of the salient features of this digitizer is that it can perform online PSD. Based on Fig.~\ref{fig3b}, a trigger signal was generated with PSD $>$ 0.2. This trigger signal was then fed to another digitizer so called "CALI" box developed at Saclay \cite{CALIbox} and used to record SPC signals with a sampling rate of 1\,{\mega\hertz}. Because the liquid scintillator pulses are fast, of the order of 10\,{\nano\second}, these pulses were slowed with an RC filter with a cutoff at 0.2 MHz and recorded by the "CALI" box in a second channel. A typical pulse from the sphere is shown in Fig.\,\ref{fig4a}. These pulses are then analyzed offline to extract various important parameters.

\section{Pulse analysis}

\begin{figure*}[]
    \centering
    \begin{subfigure}[b]{0.95\textwidth}
        \centering
        \includegraphics[width=0.95\linewidth]{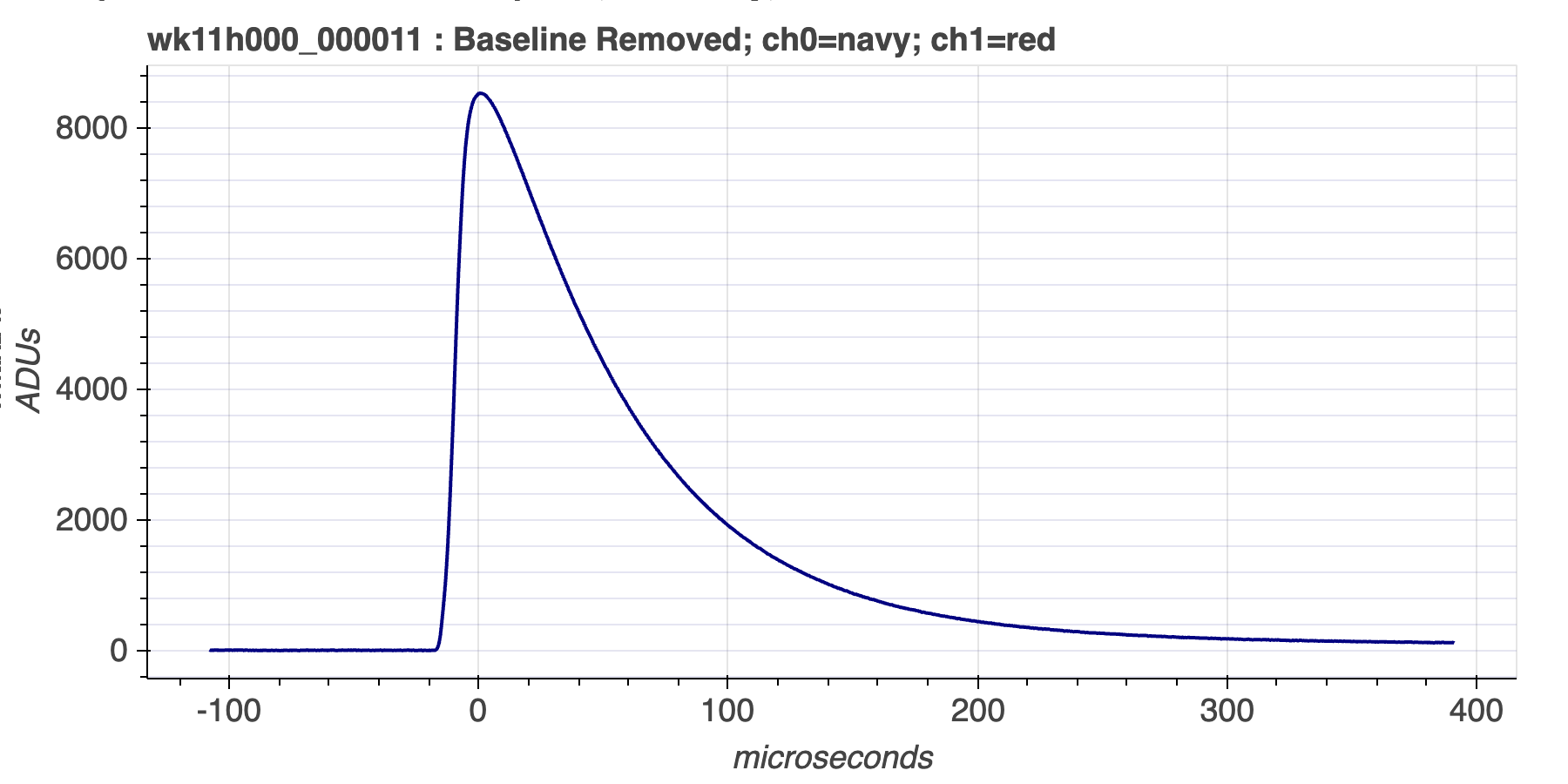}
        
        \caption{}
        \label{fig4a}
    \end{subfigure}%
  \\
    \begin{subfigure}[b]{0.9\textwidth}
        \centering
        \includegraphics[width=0.9\linewidth]{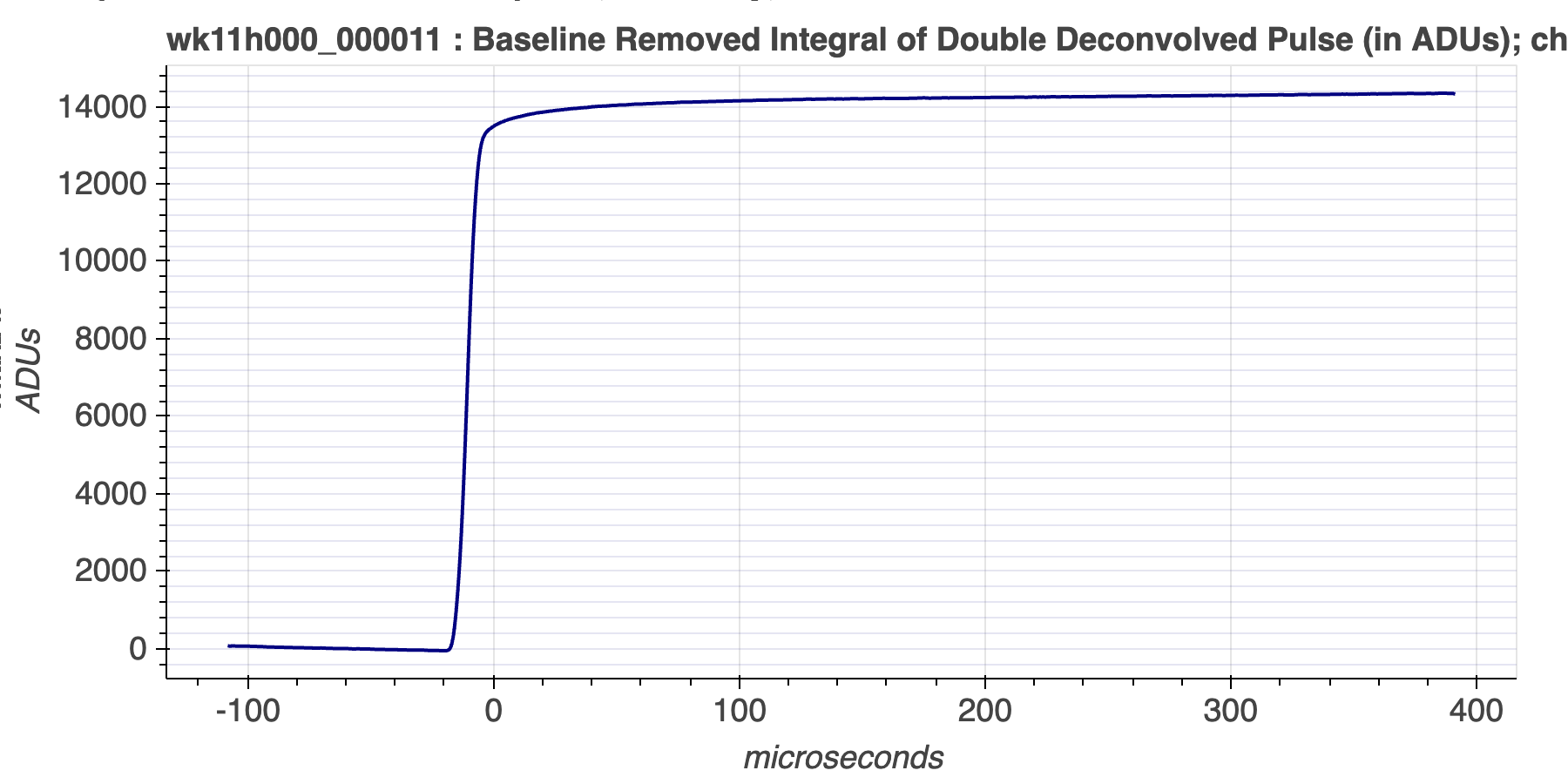}
        
        \caption{}
        \label{fig4b}
    \end{subfigure}
    \caption{A typical spherical proportional counter pulse. (a) A raw pulse with subtracted baseline. (b)  Integrated deconvolved pulse.}
     \label{fig4:typical_event}
\end{figure*}

A detailed description of the pulse analysis and its underlying principles can be found in \cite{pacothesis}. In a nutshell, the untreated signal or raw pulse undergoes baseline subtraction as shown in \ref{fig4a}, followed by smoothing with a trapezoidal filter. It is then double deconvolved using a Fourier deconvolution method to account for the preamplifier response and shape of the ion-induced current. This process generates a signal with a shape resembling a Dirac delta function, which corresponds to the arrival time of the primary electrons. Finally, the pulse is integrated, as shown in Fig. \ref{fig4b}. 

The amplitude of the event is determined by the height of the integrated pulse and is directly proportional to the energy of the interaction. The rise time is defined as the time interval for the pulse to rise from 10$\%$ to 90$\%$ of its maximum height and is a measure of the radial distribution of the incident particle's energy. The time taken for primary electrons to reach the sensor is defined as the drift time, which indicates the radial position of the event within the sphere. Assuming the start of the liquid scintillator pulse as the start of the event in the sphere, the drift time is measured as the time difference between the trigger pulse and 10$\%$ of the rise time.

\section{Event discrimination}
\label{sec:pt_tl}
 
Due to its geometry and electric field, the drift and diffusion characteristics of an SPC caused by a point-like interaction are different than those of track-like ones. For point-like events, the electrons diffuse as they move towards the anode. The more the electrons diffuse, the wider the pulse, resulting in a larger rise time. The electrons diffuse more (less) if the event happens farther from (closer to) the anode. Thus for point-like events, the rise time is an indicator of the diffusion time of the electrons in the sphere. The drift time is proportional to the radial position of the event within the sphere. As a result, in the case of point-like events, where the energy deposition is concentrated in a small region, the drift time is positively correlated with the rise time of the pulse. 
\begin{figure}[]
\centering
\includegraphics[width=0.5\textwidth]{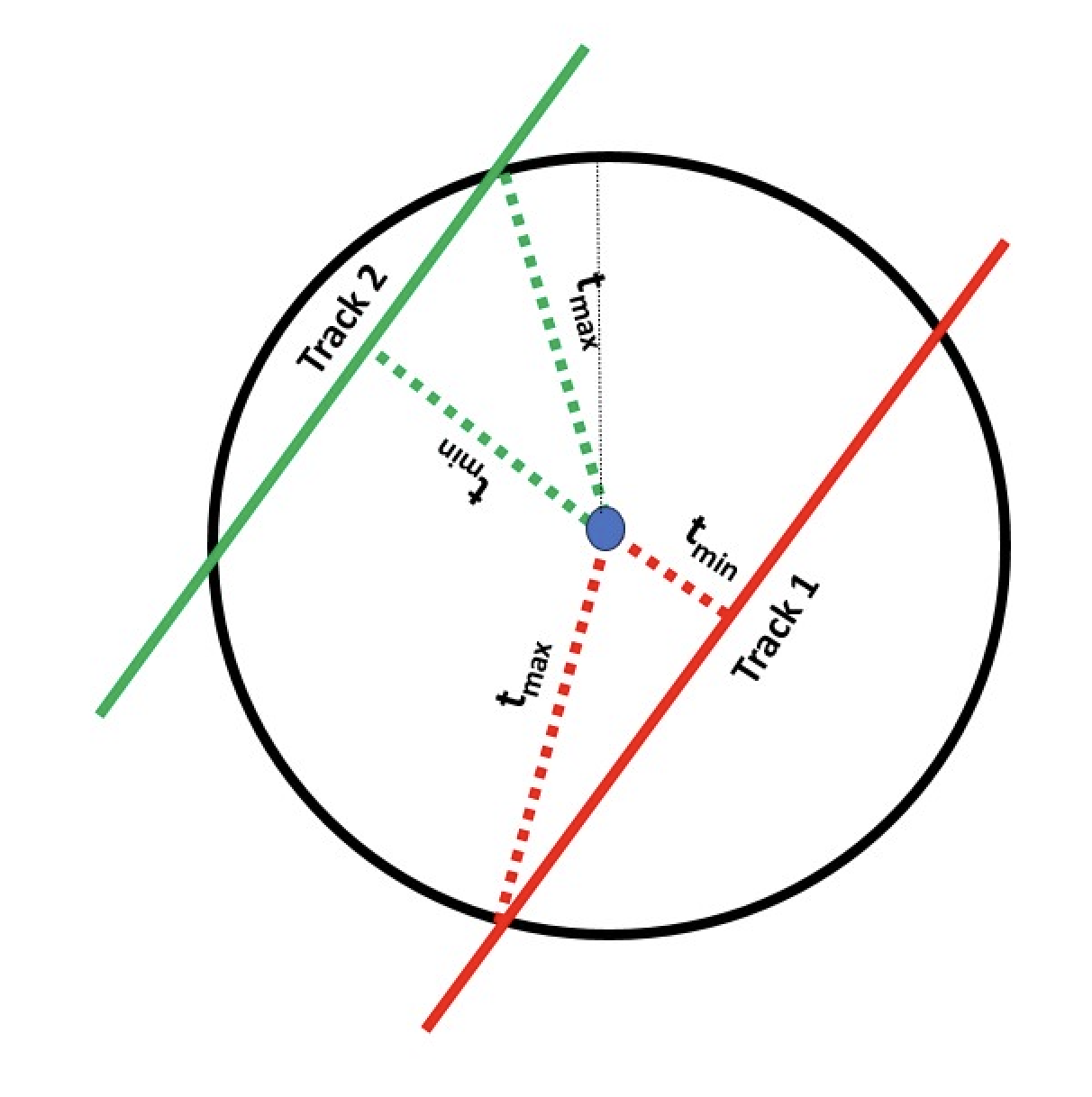}
\caption{Track-like events.}
\label{fig51:track_like_anti}
\end{figure}

In the case of track-like events, the charged particles deposit energy along the length of the track which is enclosed within the detector. The rise time is then no longer dominated by the diffusion time of the electrons from a single point within the detector volume. The pulse is a result of accumulation of all the primary electrons coming from all over the track to the anode. Therefore, the rise time represents the difference in the arrival time between the parts of the track which are farther and closer to the anode. Hence, the rise time for track-like events is higher than that of point-like events. In other words, if the arrival time of electrons closer to and farther from the anode are denoted by t$_{min}$ and t$_{max}$ respectively, then t$_{rise}$ is proportional to t$_{max}$ - t$_{min}$ and drift time t$_{drift}$ is proportional to t$_{min}$. It is interesting to note that for two tracks, one closer to the anode than the other (tracks 1 and 2, respectively) as depicted in Fig.~\ref{fig51:track_like_anti}, the drift and rise times are inversely related. This is because t$_{min, 1}$ $<$ t$_{min, 2}$, therefore t$_{drift, 1}$ $<$ t$_{drift, 2}$. However, t$_{max, 1}$ - t$_{min, 1}$ $>$ t$_{max, 2}$ - t$_{min, 2}$, therefore t$_{rise, 1}$ $>$ t$_{rise, 2}$. 

The primary ionization produced by nuclear recoil events is localized. Whereas a high-energy muon $\gamma$ generates primary electrons along the path of the particles, leaving a track. The ambient $\gamma$s produces electrons predominantly via compton scattering, mostly in the metal of the sphere, which again leaves a track. We utilized this technique to distinguish neutron events from a major source of background caused by cosmic muons and $\gamma$ events arising due to ambient radioactivity. 

\section{Neutron scattering studies}

\begin{figure*}[]
    \centering
    \begin{subfigure}[b]{0.9\textwidth}
        \centering
        \includegraphics[width=0.9\linewidth]{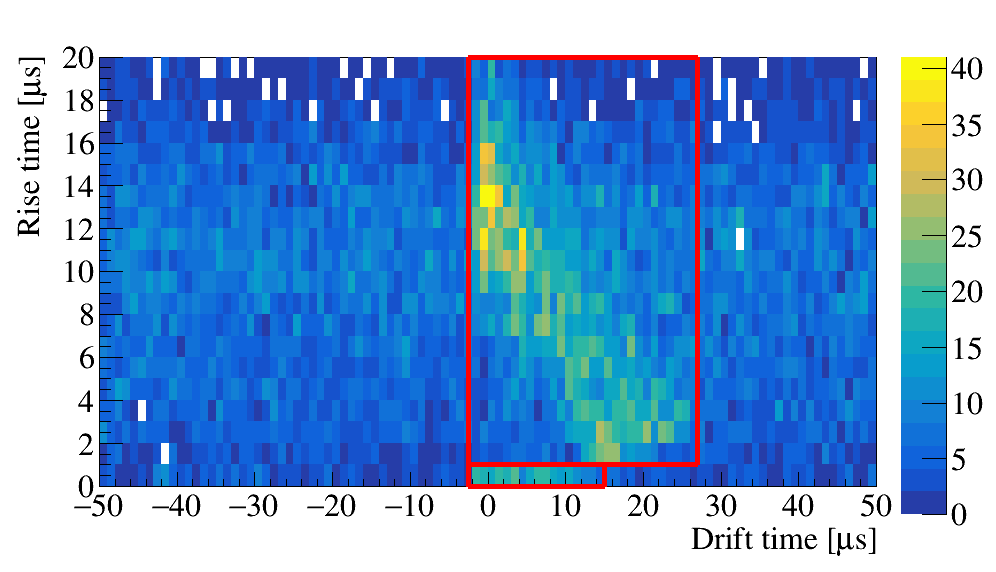}
        
        \caption{}
        \label{fig5a}
    \end{subfigure}%
  \\
    \begin{subfigure}[b]{0.95\textwidth}
        \centering
        \includegraphics[width=0.95\linewidth]{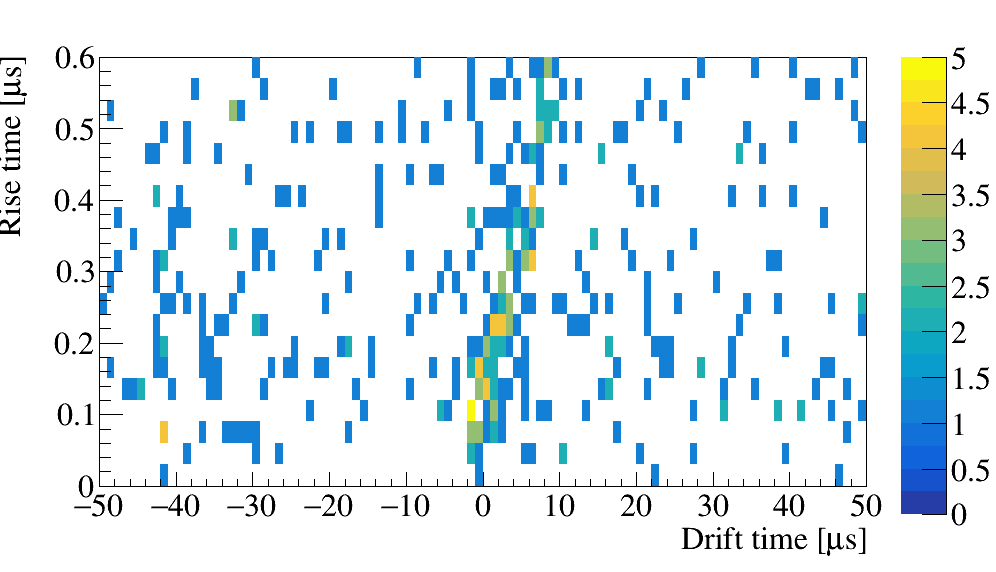}
        
        \caption{}
        \label{fig5b}
    \end{subfigure}
    \caption{Rise time vs drift time plot for a gas mixture of Ar + CH4 (2\%) at 1 bar pressure in a 15 cm diameter spherical proportional counter. ($\textit{a}$) Note the two distinct regions, showing track- and point-like events, with important relationships highlighted in red. ($\textit{b}$) To show the positive co-relation between rise time and drift time, only last y-bin is plotted.}
     \label{fig5:rtdt_s15}
\end{figure*}

The variations in the rise time-drift time relationship were used to study the effects of different gas mixtures, operating voltages and pressures on the detector characteristics. The first test was done with a mixture of Ar + 2\% CH$_{4}$ at a pressure of 1 bar in S15, giving the drift time vs rise time plotted in Fig. \ref{fig5a}, which has two distinct regions for track-like and point-like events.

\begin{figure*}[ht!]
    \centering
    \begin{subfigure}[b]{0.9\textwidth}
        \centering
        \includegraphics[width=0.9\linewidth]{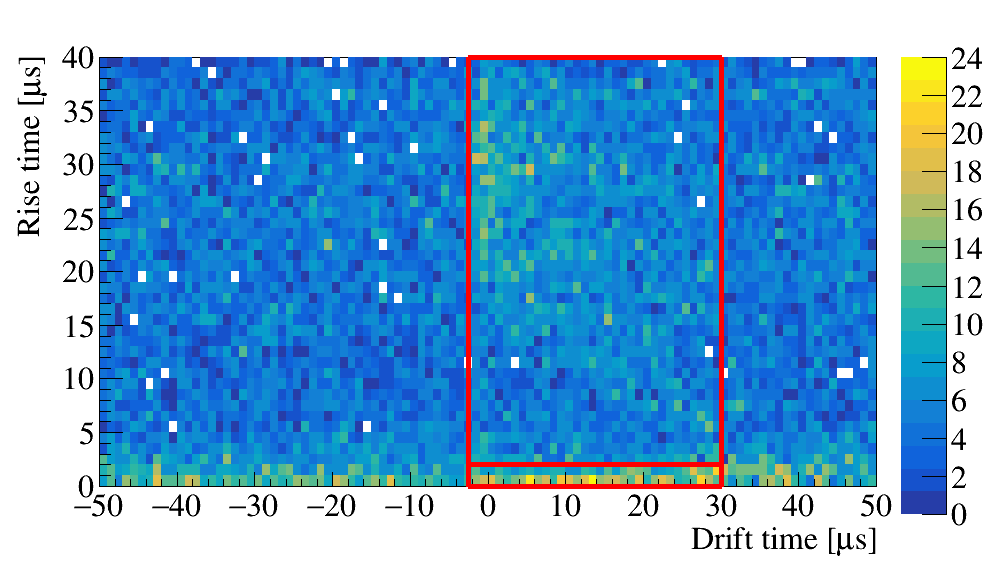}
        
        \caption{}
        \label{fig6a}
    \end{subfigure}%
  \\
    \begin{subfigure}[b]{0.9\textwidth}
        \centering
        \includegraphics[width=0.9\linewidth]{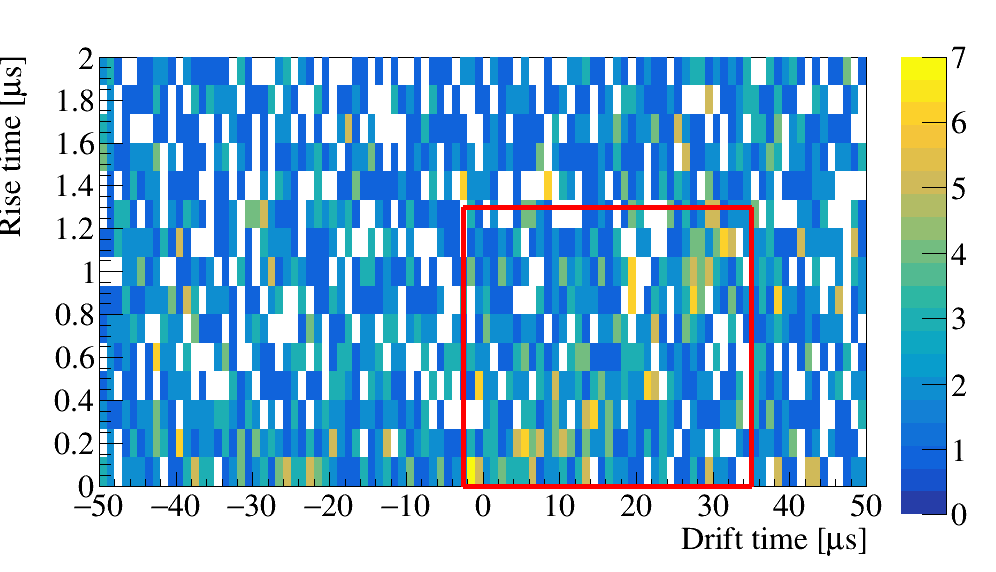}
        
        \caption{}
        \label{fig6b}
    \end{subfigure}
    \caption{Rise time vs drift time plot for a gas mixture of Ar + 2\% CH$_{4}$ at a pressure of 1 bar in a 30 cm diameter spherical proportional counter. ($\textit{a}$) Note the two distinct regions, showing track- and point-like events, with important relationships highlighted in red. ($\textit{b}$) To show the positive co-relation between rise time and drift time, only last y-bin is plotted.}
     \label{fig6:rtdt_s30}
\end{figure*}

In the point-like events region, there is a positive correlation from neutrons whereas in the track events region there is an anti-correlation between drift time and rise time due to gamma rays; these relationships are highlighted in red. The Am-Be source emits a 4.4\,{\mega}{\electronvolt} $\gamma$-ray associated with the emission of each neutron \cite{PAL1998475}. Thus, the characteristic feature of the track events discussed in section \ref{sec:pt_tl} can be seen in Fig.\,\ref{fig5a}. A similar test was done with a 30\,{\centi\meter} sphere (S30), and the results are shown in Fig.~\ref{fig6a}. 
However, a large background was observed, caused by random coincidences between neutron scattering in the larger gas volume and uncorrelated signal in the liquid scintillator. This is an irreducible background without increasing the size or number of the backing detector. Due to this, only S15 was used in the subsequent tests. 

\begin{figure}[]
\centering
\includegraphics[width=0.9\textwidth, height=0.6\textwidth]{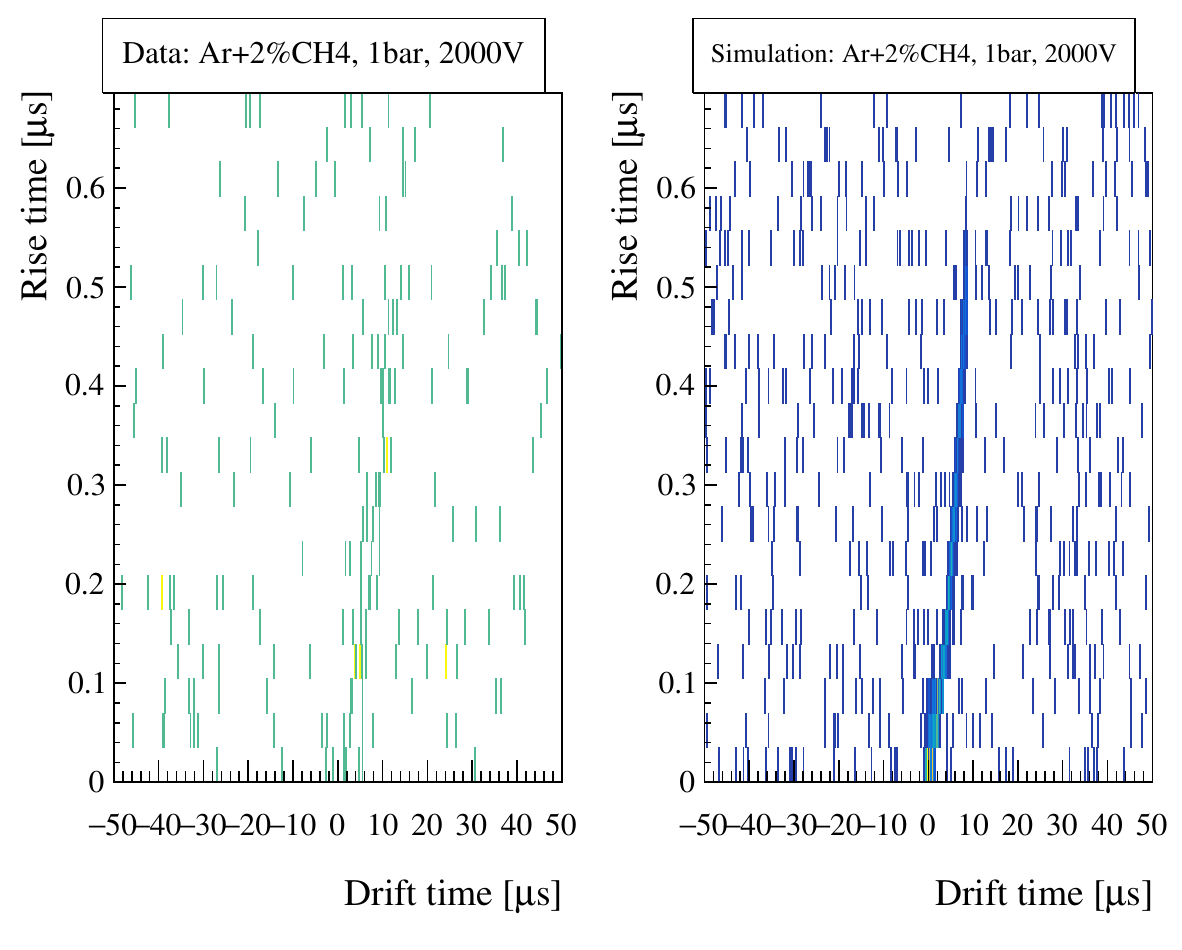}
\caption{Rise time vs drift time plot for a gas mixture of Ar + 2\% CH$_{4}$ at a pressure of 1\,bar in a 15 cm diameter spherical proportional counter, with experimental (left) and Magboltz simulation (right), results.}
\label{fig7:c_Ar_1bar}
\end{figure}

The drift and diffusion characteristics were also made using parameters computed with a simple Magboltz~\cite{magboltz} simulation to compare with the experimental result. An ideal spherical electric field $V/r^2$ from central anode in SPC was assumed in the simulation. Induced charge based on the drift of positive ions in the ideal field was taken into account in the simulation. The amplification due to the avalanche was made following a Polya distribution. Finally, the exponential response of the amplifier were convoluted with the simulation output and a normalised Gaussian noise was introduced. This was done to produce simulation output in the same format as that of the experimentally obtained data so both experimental and simulation data could be treated in the same analysis chain. The drift vs diffusion time characteristics obtained from this simulation for Ar + 2$\%$ CH$_{4}$ at a pressure of 1\,bar is shown in Fig. \ref{fig7:c_Ar_1bar} in comparison with the experimental result.

\begin{figure}[]
\centering
\includegraphics[width=0.9\textwidth]{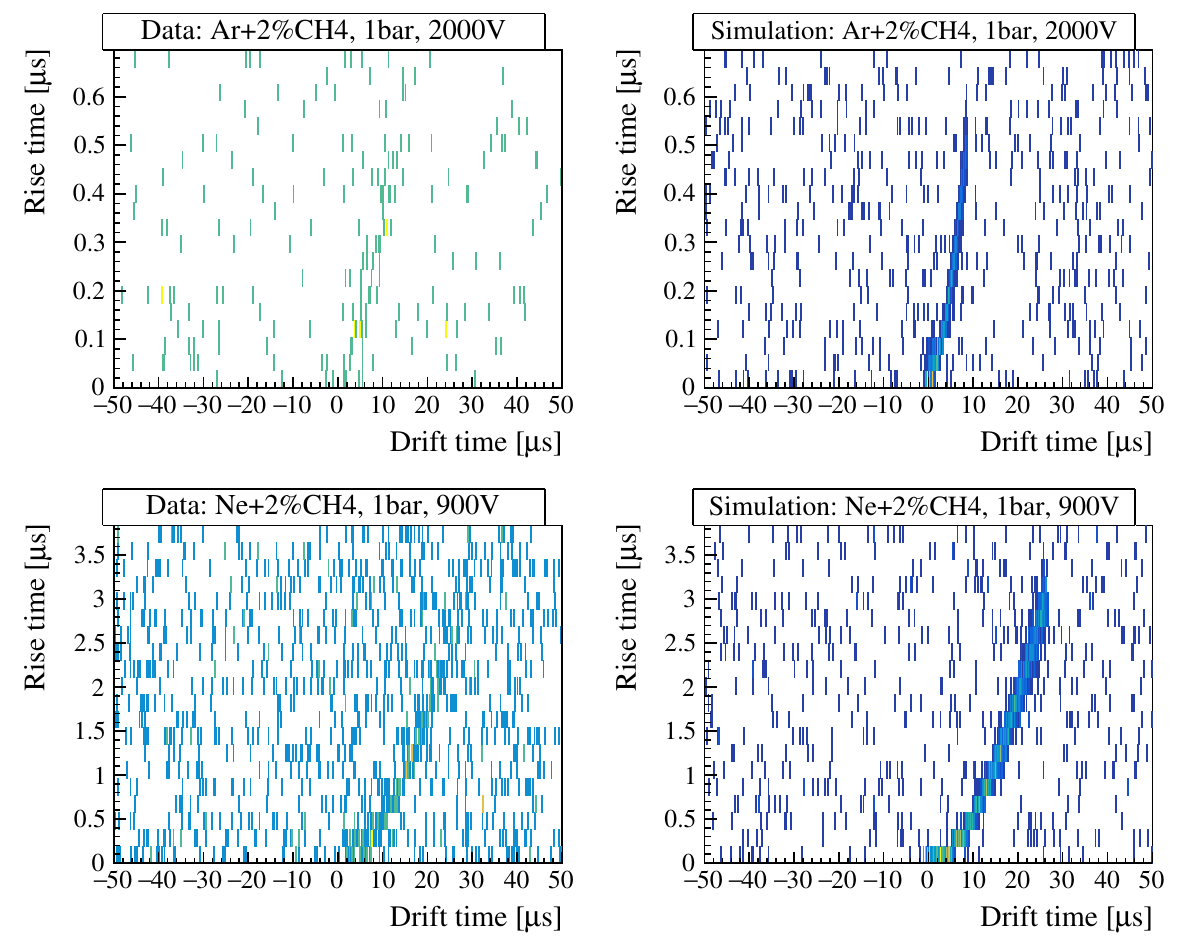}
\caption{Comparison of rise time vs drift time for Ar + 2\% CH$_{4}$ and Ne + 2\% CH$_{4}$ gas mixtures at a pressure of 1\,bar in a 15 cm diameter spherical proportional counter. Magboltz simulation and experimental results are shown.}
\label{fig8:rtdt_diff_gas}
\end{figure}

A number of tests were done to establish a proof of principle for this study. After Ar + 2$\%$ CH$_{4}$, the S15 was tested with a mixture of Ne + 2$\%$ CH$_{4}$ at a pressure of 1\,bar. Because electrons move slower in the neon mixture than in the argon one, a larger drift time was observed in the neon gas mixture. This is due to the smaller atomic number of neon leading to a higher average velocity of the gas molecules, which results in an increase of the collision probability, hence reducing the electron mean free path in the gas, as shown in Fig. \ref{fig8:rtdt_diff_gas}.

\begin{figure}[]
\centering
\includegraphics[width=0.9\textwidth]{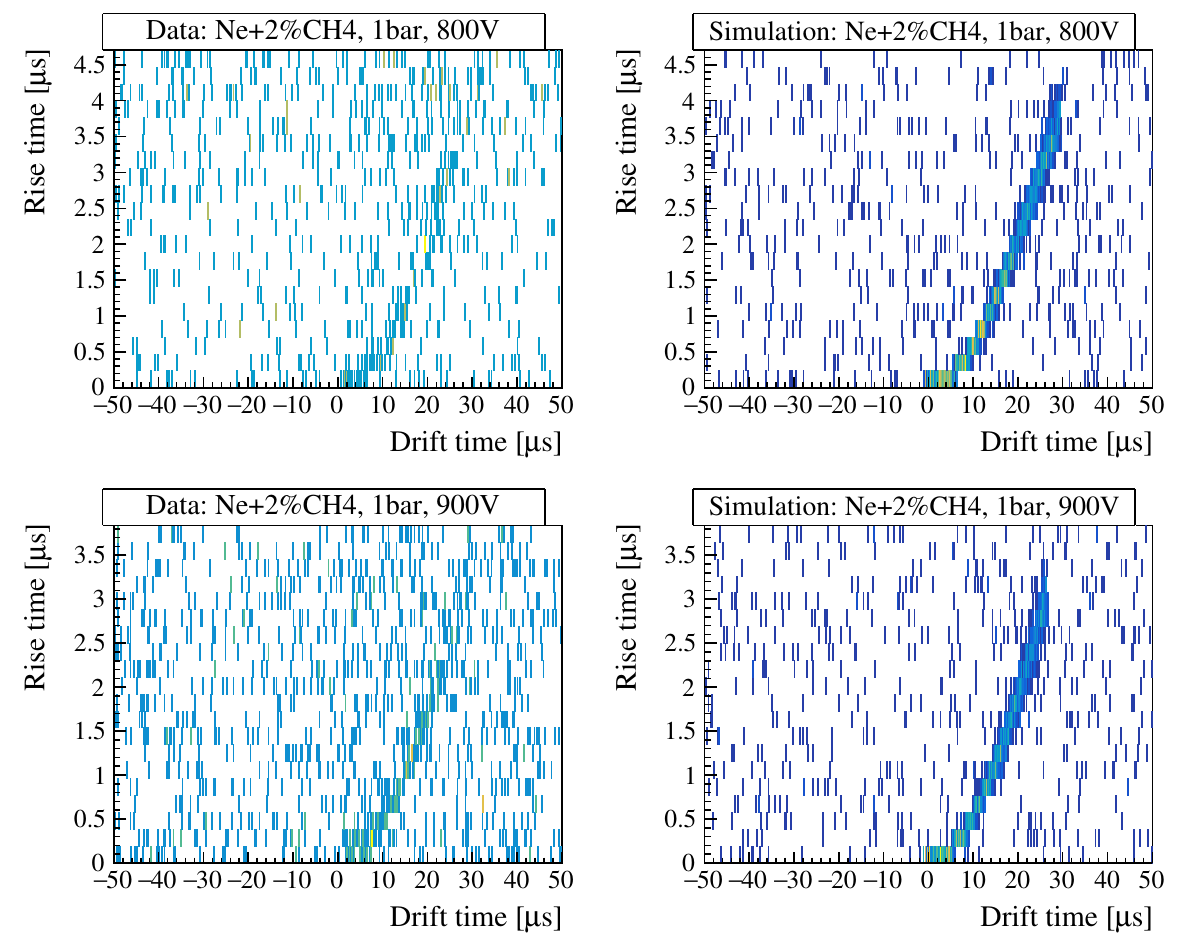}
\caption{Comparison of the rise time vs drift time is shown for a gas mixture of Ne + 2\% CH$_{4}$ at a pressure of 1\,bar in a 15 cm diameter spherical proportional counter at operating voltages of 800 and 900\,\volt. Magboltz simulation and experimental results are shown.}
\label{fig9:rtdt_diff_HV}
\end{figure}

\begin{figure}[]
\centering
\includegraphics[width=0.9\textwidth]{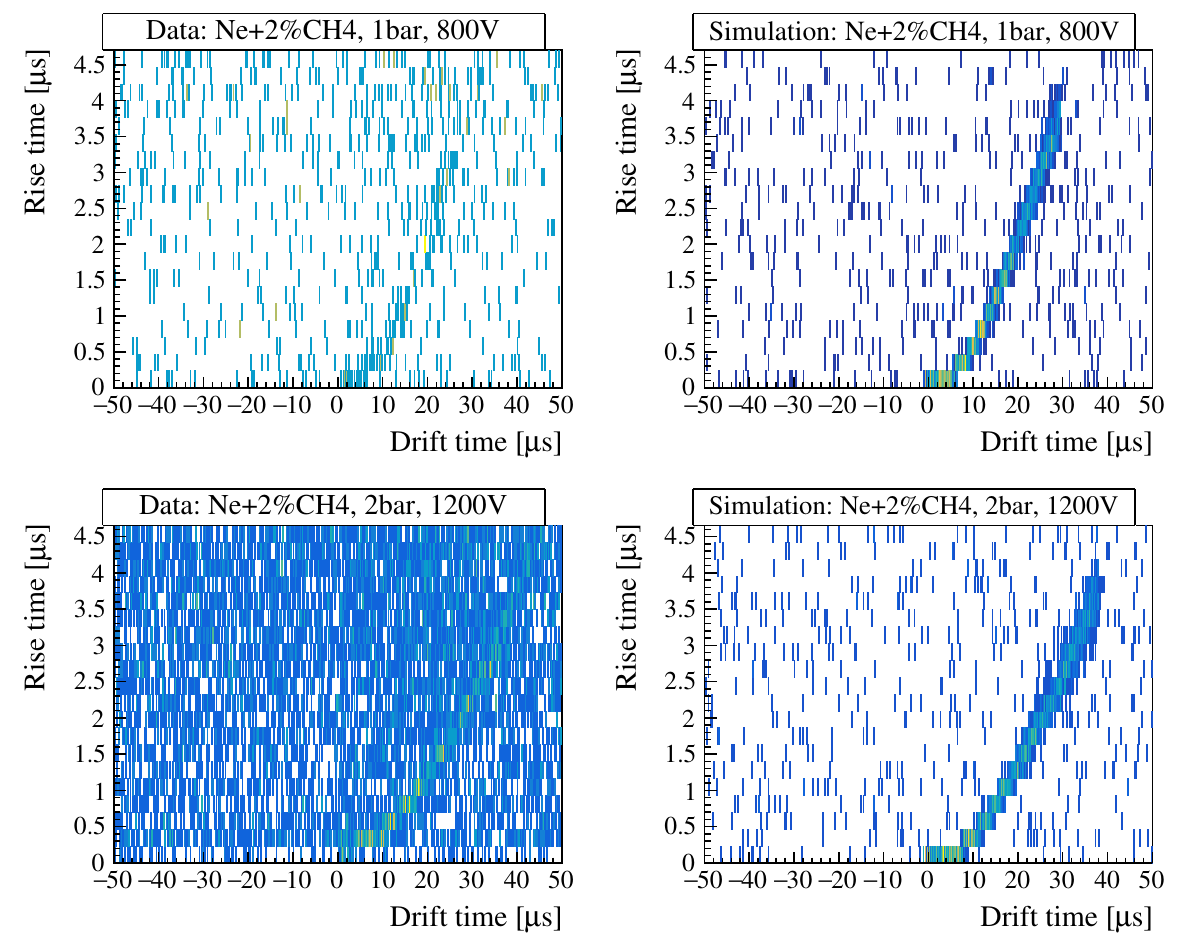}
\caption{Comparison of the rise time vs drift time is shown for a gas mixture of Ne + CH4 (2\%) in a 15 cm diameter spherical proportional counter at pressures of 1 and 2\,bar. Magboltz simulation and experimental results are shown.}
\label{fig10:rtdt_diff_P}
\end{figure}

\begin{figure}[ht!]
\centering
\includegraphics[width=0.9\textwidth, height=0.9\textwidth]{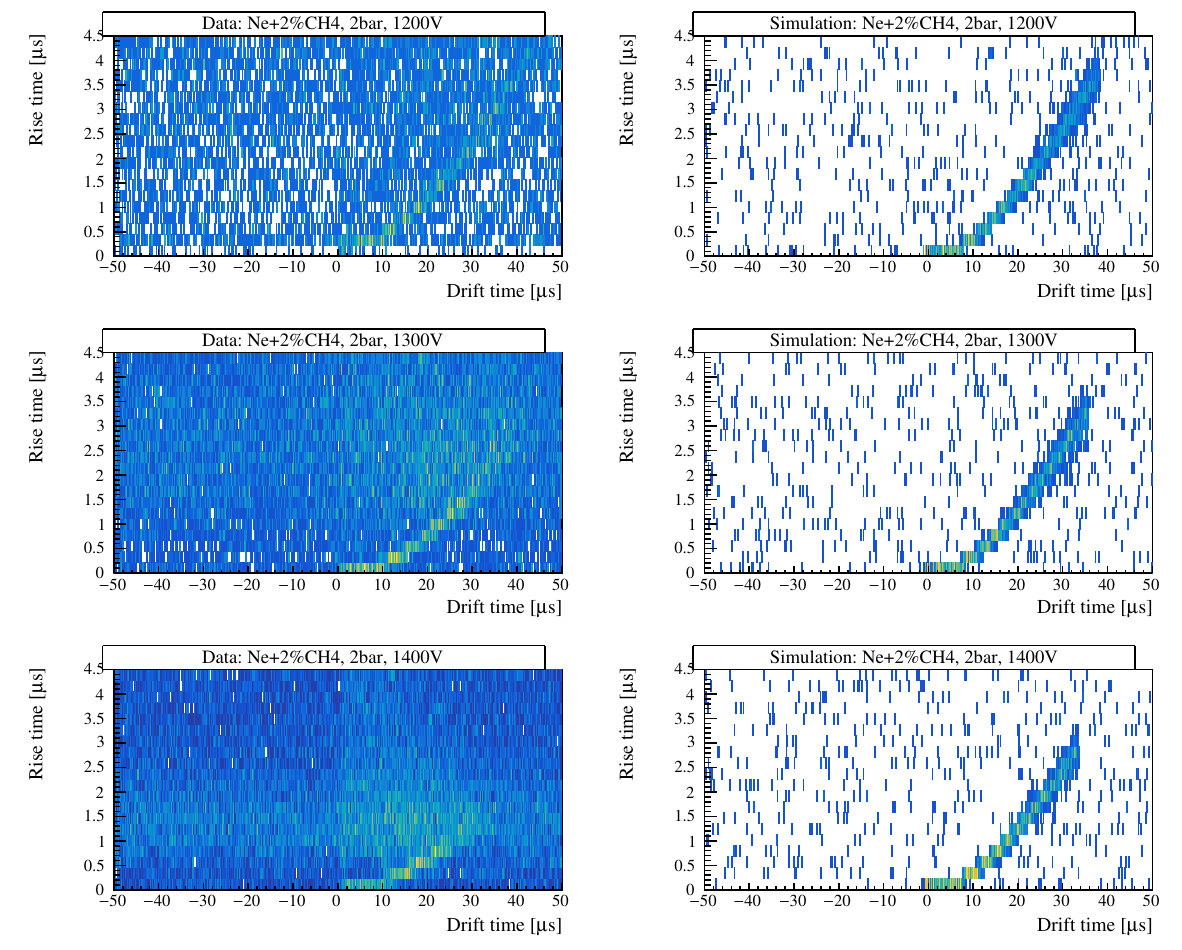}
\caption{Comparison of the rise time vs drift time is shown for a gas mixture of Ne + 2\% CH$_{4}$ at a pressure of 2\,bar in a 15 cm diameter spherical proportional counter at operating voltages of 1200, 1300 and 1400\,\volt. Magboltz simulation and experimental results are shown.}
\label{fig11:rtdt_diff_HVP}
\end{figure}

The effect of increasing the operating voltage on the drift time was also studied. The plots for Ne + 2\% CH$_{4}$ mixture at a pressure 1 bar at 800\,{\volt} and 900\,{\volt} is shown in Fig.~\ref{fig9:rtdt_diff_HV}. An increase in the electric field leads to a decrease in drift time, as shown in Fig.~\ref{fig9:rtdt_diff_HV}.
The effect of increasing the pressure on the drift time was also studied. The plots for the Ne + 2$\%$ CH$_{4}$ mixture at a pressure of 1\,bar and 2\,bar are shown in Fig. \ref{fig10:rtdt_diff_P}. An increase in the pressure leads to an increase in the drift time, as shown in Fig. \ref{fig10:rtdt_diff_P}.
The last set of measurements were done with a gas mixture of Ne + 2\% CH$_{4}$ at a pressure of 2\,bar and the operating voltage was varied as shown in Fig. \ref{fig11:rtdt_diff_HVP}.

\section{Conclusion}

This study successfully observed neutron scattering in gases through a tabletop experiment employing an AmBe source for the first time. The scattering was extensively studied with different gas mixtures, pressures, operating voltages, and sphere sizes and our findings are in good agreement with MagBoltz simulations. The tools and techniques that were developed and used for this experiment would be directly adopted for the in-beam QF measurement experiment. Therefore, it facilitates the final goal of determining the quenching factor for the NEWS-G experiment using a mono-energetic neutron beam. The systematic studies which were carried out in this experiment using different detector parameters showcases a possibility of using this technique as a potential characterization tool for new gases and sensor designs in the future.

 \bibliographystyle{elsarticle-num} 
 \bibliography{main}

\end{document}